\documentclass[reprint,showpacs,preprintnumbers,amsmath,amssymb,longbibliography,aps,pra]{revtex4-1}
\usepackage[utf8]{inputenc}
\usepackage{mathrsfs}
\usepackage{amsmath}
\usepackage{bm}
\usepackage{amsfonts}
\usepackage{amssymb}
\usepackage{amsthm}
\usepackage{color}
\usepackage{graphicx}
\graphicspath{{./images/}}
\usepackage{geometry}
\usepackage{hyperref}
\usepackage{ulem}
\usepackage{epstopdf}
\usepackage{graphicx}
\hypersetup{
     unicode=false,
     pdftoolbar=true,
     pdfmenubar=true,
     pdffitwindow=false,     % window fit to page when opened
     pdfstartview={FitH},    % fits the width of the page to the window
     pdftitle={My title},    % title
     pdfauthor={Author},     % author
     pdfsubject={Subject},   % subject of the document
     pdfcreator={Creator},   % creator of the document
     pdfproducer={Producer}, % producer of the document
     pdfkeywords={keyword1} {key2} {key3}, % list of keywords
     pdfnewwindow=true,      % links in new PDF window
     colorlinks=true,       % false: boxed links; true: colored links
     linkcolor=red,          % color of internal links (change box color with linkbordercolor)
     citecolor=blue,        % color of links to bibliography
     filecolor=magenta,      % color of file links
     urlcolor=blue           % color of external links
}
\usepackage[toc,page]{appendix}

\begin{document}
\title{Probing the Dzyaloshinskii-Moriya interaction via the propagation of spin waves in ferromagnetic thin films}
\author{Zhenyu Wang}
\author{Beining Zhang}
\author{Yunshan Cao}
\author{Peng Yan}
\email[Corresponding author: ]{yan@uestc.edu.cn}
\affiliation{School of Electronic Science and Engineering and State Key Laboratory of Electronic Thin Film and Integrated Devices, University of Electronic Science and Technology of China, Chengdu 610054, China}

\begin{abstract}
The Dzyaloshinskii-Moriya interaction (DMI) has attracted considerable recent attention owing to the intriguing physics behind and the fundamental role it played in stabilizing magnetic solitons, such as magnetic skyrmions and chiral domain walls. A number of experimental efforts have been devoted to probe the DMI, among which the most popular method is the Brillouin light scattering spectroscopy (BLS) to measure the frequency difference of spin waves with opposite wave vectors $\pm\mathbf{k}$ perpendicular to the in-plane magnetization $\mathbf{m}$. Such a technique, however, is not applicable for the cases of $\mathbf{k}\parallel\mathbf{m}$, since the spin-wave reciprocity is recovered then. For a narrow magnetic strip, it is also difficult to measure the DMI strength using BLS because of the spatial resolution limit of lights. To fill these gaps, we propose to probe the DMI via the propagation of spin waves in ferromagnetic films. We show that the DMI can cause the non-collinearity of the group velocities of spin waves with $\pm\mathbf{k}\parallel\mathbf{m}$. In heterogeneous magnetic thin films with different DMIs, negative refractions of spin waves emerge at the interface under proper conditions. These findings enable us to quantify the DMI strength by measuring the angle between the two spin-wave beams with $\pm\mathbf{k}\parallel\mathbf{m}$ in homogeneous film and by measuring the incident and negative refraction angles in heterogeneous films. For a narrow magnetic strip, we propose a nonlocal scheme to determine the DMI strength via nonlinear three-magnon processes. We implement theoretical calculations and micromagnetic simulations to verify our ideas. The results presented here are helpful for future measurement of the DMI and for designing novel spin-wave spintronic devices.
\end{abstract}

\maketitle
\section{Introduction}
The Dzyaloshinskii-Moriya interaction (DMI) is the antisymmetry component of exchange couplings, which was initially proposed to explain the weak ferromagnetism of antiferromagnets \cite{Dzyaloshinsky1958,Moriya1960}. This interaction originates from the spin-orbit coupling in magnetic materials with broken inversion symmetry, either in bulk or at the interface. Recently, the DMI has drawn extensive research interest due to two main reasons: (i) its fundamental role in stabilizing topological magnetic solitons, such as skyrmions \cite{Mhlbauer2009,Yu2010,Sampaio2013,Woo2016,Boulle2016} and chiral domain walls \cite{Heide2008,Chen201304,Chen2013,Benitez2015,Tetienne2015}, which are promising candidates for future spintronic applications; (ii) the intriguing physics associated with the nonreciprocal propagation of spin waves (magnons) \cite{Lan2015,Xing2016,Garcia2014,Bracher2017}, the elementary excitations in ordered magnets. The determination of the DMI is thus an important issue.

Several experimental schemes have been proposed to measure the DMI strength. For example, it can be quantified by imaging the profile of chiral domain walls \cite{Heide2008,Chen201304,Chen2013} or by analyzing their dynamical behaviors \cite{Benitez2015,Hiramatsu2014,Kim2017,Balk2017,Soucaille2016} when the driving electric currents and/or magnetic fields are applied. When the DMI is not strong enough to stabilize the inhomogeneous magnetic texture (such as the domain wall), the spin-wave excitation carries the unique information of the DMI. Recent experiments have demonstrated that the DMI constant can be determined by measuring the frequency difference $[\Delta\omega=\omega(\mathbf{k})-\omega(-\mathbf{k})]$ of spin waves with opposite wave vectors ($\pm\mathbf{k}$) perpendicular to the magnetization ($\mathbf{m}$) using the Brillouin light scattering spectroscopy (BLS) \cite{Soucaille2016,Belmeguenai2015,Cho2015,Di2015,Nembach2015,Hrabec2017}, the spin-polarized electron energy loss spectroscopy \cite{Zakeri2010}, and the propagating spin wave spectroscopy \cite{Lee2016}. For the case of $\mathbf{k}\parallel\mathbf{m}$, the frequency difference of spin waves with $\pm\mathbf{k}$ vanishes and these schemes are unfeasible. We note that spin-wave excitations in these experiments are in the long wavelength regime, where the anisotropic dipolar interaction cannot be ignored. The nonreciprocal nature of dipolar interactions may blur the quantification of the DMI. Moreover, for a magnetic strip with the width well below 100 nm, it is difficult to utilize the BLS to measure the DMI owing to the diffraction limit of lights. It is, therefore, necessary to develop new methods to measure the DMI for the situations mentioned above.

In this work, we propose to probe the DMI strength in ferromagnetic films via the propagation of spin waves. To this end, we systematically investigate the effects of the DMI on the propagation, the scattering, and the interaction of spin waves in different magnetic structures. We first consider a homogeneous ferromagnetic film, and find that the DMI induces a non-collinearity between the wave vector and group velocity ($\mathbf{v}_{\mathrm{g}}=\mathrm{d}\omega/\mathrm{d}\mathbf{k}$) of spin waves when $\mathbf{k}$ is not perpendicular to $\mathbf{m}$. Spin-wave canting induced by the DMI has been reported in ferromagnetic nanowires \cite{Guo2017} with $\mathbf{v}_{\mathrm{g}}\parallel\mathbf{m}$. Here, we predict another non-collinearity between two spin-wave beams with opposite wave vectors $\pm\mathbf{k}\parallel\mathbf{m}$. The angle between the two spin-wave beams is derived analytically. Since this non-collinearity comes from the DMI but not the dipolar interaction, we can exclusively determine the DMI strength by measuring the angle between the two beams. Inspired by recent advances of spatially modulated DMI in heterogeneous ferromagnetic films \cite{Mulkers2017,Lee2017,Hong2017}, we then investigate the spin-wave scattering at the interface separating two co-planar ferromagnets with different DMIs. We focus on the exchange spin-wave region, where the nonlocal dipolar effect can be approximated by local demagnetizing fields. We obtain the generalized Snell's law, and show the emergence of both the negative refraction and the total reflection under proper spin-wave incident angles. These peculiar phenomena and the generalized Snell's formula can be used to quantify the DMI strength by simply measuring the incident and refracted angles of spin-wave beams, which can be readily realized by direct imagings \cite{Stigloher2016}.

Recently, we developed a three-magnon interaction method to detect spin waves localized in the magnetic domain wall nanochannels \cite{Zhang2018}. The approach can be parallelly applied for probing the DMI in narrow magnetic strips. In general, the three-magnon process is triggered by the weak nonlocal magnetic dipole-dipole interaction in uniform ferromagnets \cite{Costa2000}. It can also occur in magnetic textures such as skyrmions \cite{Aristov2016} and domain walls \cite{Zhang2018} without the dipolar interaction. Here, we consider another three-magnon effect induced by the DMI in uniform ferromagnets. The idea is analytically formulated with micromagnetic simulations performed to verify the theoretical predictions. All micromagnetic simulations in this work are performed using the OOMMF package \cite{Donahue1999,Rohart2013}.

The structure of this paper is organized as follows. In Sec. II, we derive the dispersion relation of spin waves in a chiral magnetic film. The group-velocity non-collinearity of two propagating spin-wave beams with anti-parallel wave vectors is presented. We also investigate the spin-wave scattering across the interface of two ferromagnets with different DMIs. In Sec. III, the three-magnon processes arising in a narrow magnetic strip with the DMI are studied. We demonstrate that this nonlinear effect can be utilized to accurately quantify the DMI constant in the strip. Conclusions are drawn in Sec. IV.

\section{The linear dynamics of spin waves in homogeneous and heterogeneous chiral magnetic films}
We first consider the spin-wave propagation in a magnetic thin film with the interfacial DMI of the following form \cite{Bogdanov2001},
\begin{equation}\label{eq1}
\begin{split}
  \mathbf{H}_{\mathrm{DM}} &= \frac{2D}{\mu_{0}M_{s}}[\nabla{m}_{z}-(\nabla\cdot\mathbf{m})\hat{z}] \\
    &= \frac{2D}{\mu_{0}M_{s}}(\frac{\partial{m_{z}}}{\partial{x}},\frac{\partial{m_{z}}}{\partial{y}},-\frac{\partial{m_{x}}}{\partial{x}}-\frac{\partial{m_{y}}}{\partial{y}}),
\end{split}
\end{equation}
where $D$ is the DMI constant, $M_{s}$ is the saturation magnetization, and $\mathbf{m}=(m_{x},m_{y},m_{z})$ is the unit magnetization vector. The magnetization dynamics is described by the Landau-Lifshitz-Gilbert (LLG) equation,
\begin{equation}\label{eq2}
  \frac{\partial\mathbf{m}}{\partial{t}}=-\gamma\mu_{0}\mathbf{m}\times\mathbf{H}_{\mathrm{eff}}+\alpha\mathbf{m}\times\frac{\partial\mathbf{m}}{\partial{t}},
\end{equation}
\begin{figure}
  \centering
  \includegraphics[width=0.5\textwidth]{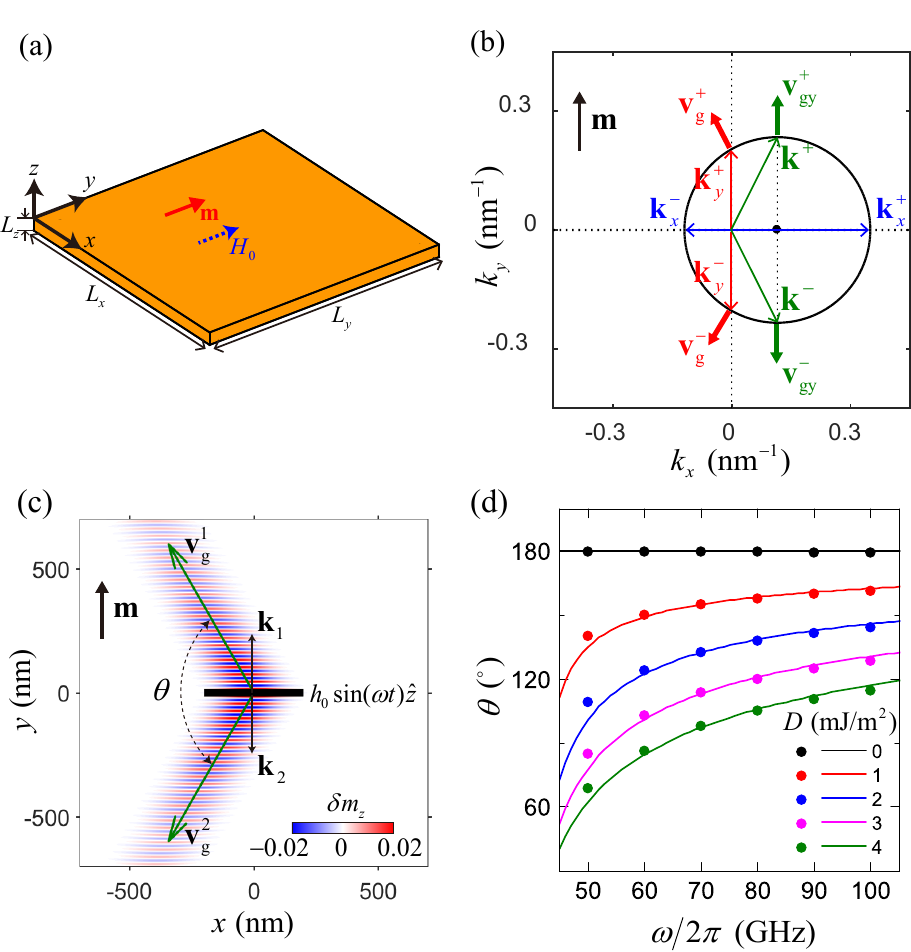}\\
  \caption{(a) Schematic illustration of an ultrathin film with width $L_{x}$, length $L_{y}$, and thickness $L_{z}$. A magnetic field $H_{0}$ is applied along $+\hat{y}$, which makes the magnetization $\mathbf{m}$ lying in the film plane ($\mathbf{m}\parallel+\hat{y}$). (b) The isofrequency curve calculated based on the dispersion relation Eq. (\ref{eq4}). $\mathbf{k}_{x}^{+}$ and $\mathbf{k}_{x}^{-}$ are the wave vectors of spin waves propagating along the direction perpendicular to the magnetization. $\mathbf{v}_{\mathrm{g}}^{+}$ and $\mathbf{v}_{\mathrm{g}}^{-}$ are the group velocities of spin waves with $\mathbf{k}_{\mathrm{y}}^{\pm}\parallel\mathbf{m}$. $\mathbf{k}^{+}$ and $\mathbf{k}^{-}$ are the wave vectors of spin waves propagating along the magnetization ($\mathbf{v}_{\mathrm{gy}}^{\pm}\parallel\mathbf{m}$). (c) Non-collinear propagation of two spin-wave beams with opposite wave vectors ($\mathbf{k}_{1}=-\mathbf{k}_{2}$). The spin-wave beams are excited by a sinusoidal monochromatic microwave source with  $\mu_{0}h_{0}=0.1$ $\mathrm{T}$ and $\omega/2\pi=80$ GHz in a rectangular region (black bar). (d) The angle between the two spin-wave beams as a function of the excitation frequency for various DMI constants $D$. The dots and curves correspond to the numerical simulation results and the analytical formula Eq. (\ref{eq6}), respectively. }\label{fig1}
\end{figure}with the gyromagnetic ratio $\gamma=1.76\times10^{11}$ $\mathrm{rad}$ $\mathrm{s^{-1}T^{-1}}$, the vacuum permeability $\mu_{0}$, and the Gilbert damping constant $\alpha$. The effective field $\mathbf{H}_{\mathrm{eff}}$ comprises of the exchange field, the DM field, the demagnetization field, and the external field. Although the DMI facilitates the inhomogeneous magnetic texture, it is possible to stabilize a single-domain structure as the ground state, when the external field is sufficiently strong. Given that the DMI has no effect on spin waves when $\mathbf{m}$ is perpendicular to the film plane \cite{Landeros2013,Garcia2014}, an external field $\mathbf{H}_{\mathrm{ext}}={H_{0}}\hat{y}$ is applied to make $\mathbf{m}$ in the film plane $(\mathbf{m}=+\hat{y})$, as shown in Fig. \ref{fig1}(a). For simplicity, the dipolar interaction is approximated by the static demagnetizing field for an extend film $\mathbf{H}_{\mathrm{d}}=-M_{s}m_{z}\hat{z}$. Neglecting the damping term ($\alpha=0$), the spin wave dispersion relation can be obtained by solving the linearized LLG equation
\cite{Landeros2013,Moon2013},
\begin{equation}\label{eq3}
  \omega(\mathbf{k})=\sqrt{(A^{\ast}\mathbf{k}^{2}+\omega_{\mathrm{H}})(A^{\ast}\mathbf{k}^{2}+\omega_{\mathrm{H}}+\omega_{\mathrm{m}})}-D^{\ast}k_{x},
\end{equation}
where $A^{\ast}=2\gamma{A}/M_{s}$ with the exchange constant $A$, $\omega_{\mathrm{H}}=\gamma\mu_{0}{H}_{0}$, $\omega_{\mathrm{m}}=\gamma\mu_{0}M_{s}$, $D^{\ast}=2\gamma{D}/M_{s}$, and $\mathbf{k}=(k_{x},k_{y})$ the wave vector of spin wave. Here we focus on the exchange spin waves with high frequencies and simplify Eq. (\ref{eq3}) to
\begin{equation}\label{eq4}
  \omega(\mathbf{k})=A^{\ast}\mathbf{k}^{2}-D^{\ast}k_{x}+\omega_{\mathrm{H}}+\frac{\omega_{\mathrm{m}}}{2},
\end{equation}
and obtain the group velocity
\begin{equation}\label{eq5}
  \mathbf{v}_{\mathrm{g}}=\frac{\partial\omega}{\partial\mathbf{k}}=(2A^{\ast}k_{x}-D^{\ast})\hat{x}+2A^{\ast}k_{y}\hat{y}.
\end{equation}

The influence of the DMI on spin-wave propagations in a chiral ferromagnetic film can be analyzed by the isofrequency curve, as shown in Fig. \ref{fig1}(b). In the presence of the DMI, the isofrequency circle shifts away from the origin in the $\mathbf{k}$ space. For $\mathbf{k}\perp\mathbf{m}$, the magnitudes of wave vectors with opposite directions are different, which indicates an asymmetry of the spin-wave wavelength with respect to the propagation direction \cite{Moon2013}. When the propagation direction ($\mathbf{v}_{\mathrm{gy}}^{+}$ and $\mathbf{v}_{\mathrm{gy}}^{-}$) of spin wave is along the magnetization, the wave vectors ($\mathbf{k}^{+}$ and $\mathbf{k}^{-}$) of spin waves become oblique with respect to the propagation direction, which is responsible for the spin-wave canting numerically observed in Ref. \cite{Guo2017}. In the case of $\mathbf{k}\parallel\mathbf{m}$ ($k_{x}=0$), the group velocity in Eq. (\ref{eq5}) can be written as $\mathbf{v}_{\mathrm{g}}=-D^{\ast}\hat{x}+2A^{\ast}k_{y}\hat{y}$. In the presence of the DMI ($D\neq0$), the group velocities ($\mathbf{v}_{\mathrm{g}}^{+}$ and $\mathbf{v}_{\mathrm{g}}^{-}$) of spin waves with opposite wave vectors ($\mathbf{k}_{\mathrm{y}}^{+}$ and $\mathbf{k}_{\mathrm{y}}^{-}$) are not collinear, as shown in Fig. \ref{fig1}(b). The angle between the two group velocities can be obtained:
\begin{align}\label{eq6}
  \theta&=\arccos\frac{\mathbf{v}_{\mathrm{g}}^{+}\cdot\mathbf{v}_{\mathrm{g}}^{-}}{|\mathbf{v}_{\mathrm{g}}^{+}||\mathbf{v}_{\mathrm{g}}^{-}|}\nonumber\\
  &=\arccos\Big[\frac{(D^{\ast})^{2}-4A^{\ast}(\omega-\omega_{\mathrm{H}}-\omega_{\mathrm{m}}/2)}{(D^{\ast})^{2}+4A^{\ast}(\omega-\omega_{\mathrm{H}}-\omega_{\mathrm{m}}/2)}\Big].
\end{align}
Based on Eq. (\ref{eq6}), the DMI strength can be evaluated by measuring the angle of the two spin-wave beams.

We confirm the above results using micromagnetic simulations. We consider a ferromagnetic thin film with length 2000 nm, width 2000 nm, and thickness 1 nm, which lies in the $x-y$ plane. Magnetic parameters of Permalloy were used in simulations: $M_{s}=8\times10^{5}$ $\mathrm{A/m}$, $A=13$ $\mathrm{pJ/m}$, and $\alpha=0.01$. In the simulations, the Gilbert damping constant close to the film edges is set to linearly increase to 1.0 to avoid the spin-wave reflection by the boundaries \cite{Berkov2006}. We apply an external field $\mu_{0}H_{0}=1$ T along $+\hat{y}$ that is sufficiently strong to saturate the magnetization in the film plane. Then, we stimulate the propagations of two spin-wave beams with opposite wave vectors parallel and antiparallel to the magnetization ($\pm\mathbf{k}\parallel\mathbf{m}$). To this end, we apply a sinusoidal monochromatic microwave source $\mathbf{H}_{\mathrm{ext}}=h_{0}\sin(\omega{t})\hat{z}$ in a narrow rectangular region ($300\times10$ $\mathrm{nm}^{2}$) [black bar shown in Fig. \ref{fig1}(c)], where the field amplitude $h_{0}$ has a Gaussian profile in the transverse direction ($\hat{x}$) \cite{Gruszecki2015}. Figure \ref{fig1}(c) shows two spin waves with $\omega/2\pi=80$ $\mathrm{GHz}$ and $\pm\mathbf{k}\parallel\mathbf{m}$ in the presence of DMI ($D=3.0$ $\mathrm{mJ/m^{2}}$). It is clear to see that the group velocities $\mathbf{v}_{\mathrm{g}}^{1}$ and $\mathbf{v}_{\mathrm{g}}^{2}$ of the two spin-wave beams are non-collinear. Both beams propagate towards the left, which is fully in line with formula $\mathbf{v}_{\mathrm{g}}=-D^{\ast}\hat{x}+2A^{\ast}k_{y}\hat{y}$. The angle $\theta=120.5^{\circ}$ obtained from simulation is also consistent with the theoretical prediction Eq. (\ref{eq6}) with a deviation less than $0.5\%$.

The angles between the two spin-wave beams measured from simulations (dots) as well as the ones obtained from Eq. (\ref{eq6}) (curves) are plotted as a function of the excitation frequencies for different DMI constants, as shown in Fig. \ref{fig1}(d). Good agreements can be found, except at low frequencies ($\omega/2\pi\leqslant50$ GHz) where the wave vectors of spin waves have a slight deviation from the presumed $\hat{y}$ direction (not shown). For $D=0$, the angle between the two spin-wave beams is always equal to $180^{\circ}$. This indicates that the non-collinearity of the two spin-wave beams observed above is exclusively induced by the DMI rather than the dipolar interaction. Thus, the DMI strength can be determined by measuring the angle between the two spin-wave beams. This method requires the spin wave imaging with the spatial resolution in the range of about 200 nm (the width of the spin-wave beam), which can be achieved by X-ray magnetic circular dichroism (XMCD) \cite{Bonetti2015} or the near-field BLS \cite{Jersch2010}.

\begin{figure}
  \centering
  \includegraphics[width=0.5\textwidth]{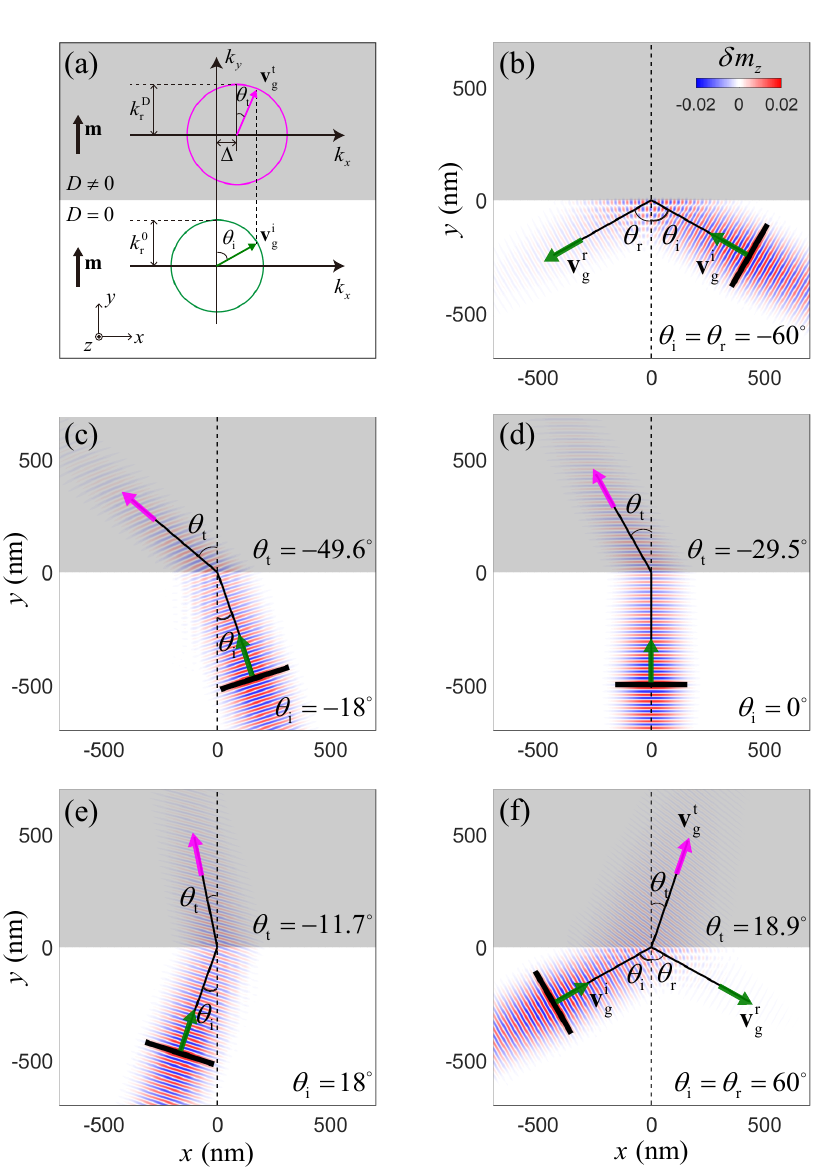}\\
  \caption{(a) Schematic plot of the generalized Snell’s law for spin-wave scattering at the DMI interface. The magnetization of ultrathin film is saturated along $+\hat{y}$ direction. The green (lower) and magenta (upper) circles correspond to the isofrequency curves in momentum space for spin-wave propagation in no-DMI and DMI regions, respectively. (b)$\sim$(f) Refraction and reflection of spin waves scattering at the DMI interface with different incident angles. (b) $\theta_{\mathrm{i}}=-60^{\circ}$, (c) $\theta_{\mathrm{i}}=-18^{\circ}$, (d) $\theta_{\mathrm{i}}=0^{\circ}$, (e) $\theta_{\mathrm{i}}=18^{\circ}$, and (f) $\theta_{\mathrm{i}}=60^{\circ}$. Arrows label the group velocities of spin waves.}\label{fig2}
\end{figure}

Next, we proceed to investigate the scattering of exchange spin waves across the interface of two ferromagnets with different DMIs. One effect of the spatially modulated DMI is the equilibrium spin canting at the DMI interface \cite{Mulkers2017}. Since spin canting only occurs at a narrow range around the DMI interface, its effect on the spin-wave propagation is rather weak. Hence, we view the magnetization of the heterogeneous film as uniform along the $+\hat{y}$ direction. Based on Eq. (\ref{eq4}), the isofrequency curves of spin waves propagation in no-DMI and DMI regions are plotted in $\mathbf{k}$ space, as shown in Fig. \ref{fig2}(a). In the absence of the DMI ($D=0$), the spin-wave isofrequency curve at a given frequency $\omega$ is a circle centered at the origin with the radius $k_{\mathrm{r}}^{0}=\sqrt{(\omega-\omega_{\mathrm{H}}-\omega_{\mathrm{m}}/2)/A^{\ast}}$. With the DMI ($D\neq0$), the isofrequency circle is shifted by $\Delta=D^{\ast}/2A^{\ast}$ along $+k_{x}$ axis and its radius increases to $k_{\mathrm{r}}^{\mathrm{D}}=\sqrt{(k_{\mathrm{r}}^{0})^{2}+\Delta^{2}}$. According to the conservation of momentum parallel with the interface, we obtain the generalized Snell’s law
\begin{equation}\label{eq7}
  k_{\mathrm{r}}^{0}\sin\theta_{\mathrm{i}}=k_{\mathrm{r}}^{\mathrm{D}}\sin\theta_{\mathrm{t}}+\Delta,
\end{equation}
where $\theta_{\mathrm{i}}$ and $\theta_{\mathrm{t}}$ are the incident angle and refraction angle of spin-wave beams with respect to the interface normal, as shown in Fig. \ref{fig2}(a). Similar results have been presented for spin waves propagation in different magnetic systems \cite{Yu2016,Mulkers2018}. For instance, the generalized Snell's law describes the spin wave refracted at the domain wall in a chiral magnetic film in Ref. \cite{Yu2016}.

The spin-wave scattering at the DMI interface can be divided into three cases: total reflection, negative refraction and normal refraction. The critical angles for total reflection ($\theta_{\mathrm{c}}^{\mathrm{T}}$) and negative refraction ($\theta_{\mathrm{c}}^{\mathrm{N}}$) are the incident angles corresponding to the refracted angles $\theta_{\mathrm{t}}=-90^{\circ}$ and $\theta_{\mathrm{t}}=0^{\circ}$, respectively. Using Eq. (\ref{eq7}), we obtain $\theta_{\mathrm{c}}^{\mathrm{T}}=\arcsin[(\Delta-k_{\mathrm{r}}^{\mathrm{D}})/k_{\mathrm{r}}^{0}]$ and $\theta_{\mathrm{c}}^{\mathrm{N}}=\arcsin[\Delta/k_{\mathrm{r}}^{0}]$. Below we examine the scattering of spin waves at the DMI interface ($D=3.0$ $\mathrm{mJ/m^{2}}$) for different incident angles. For spin waves with $\omega/2\pi=80$ GHz, the critical angles for total reflection and negative refraction are $\theta_{\mathrm{c}}^{\mathrm{T}}=-35.7^{\circ}$ and $\theta_{\mathrm{c}}^{\mathrm{N}}=34.4^{\circ}$. For incident angles $-90^{\circ}<\theta_{\mathrm{i}}<-35.7^{\circ}$, there is no real solution for the refracted angle $\theta_{\mathrm{t}}$ and total reflection takes place. This situation is plotted in Fig. \ref{fig2}(b), showing that the incident spin wave with $\theta_{\mathrm{i}}=-60^{\circ}$ cannot transmit into the DMI region (upper gray region) and is completely reflected. For $-35.7^{\circ}<\theta_{\mathrm{i}}<0^{\circ}$, the incident and refracted angles have the same sign corresponding to normal refraction, as shown in Fig. \ref{fig2}(c) with $\theta_{\mathrm{i}}=-18^{\circ}$. In the case of vertical incidence, the refracted angle is $-29.5^{\circ}$ rather than $0^{\circ}$, which is different from its optical analog, as illustrated in Fig. \ref{fig2}(d). For $0^{\circ}<\theta_{\mathrm{i}}<34.4^{\circ}$, the refracted angle is negative. As an example, we set $\theta_{\mathrm{i}}=18^{\circ}$ in Fig. \ref{fig2}(e), and find that the refracted angle is $\theta_{\mathrm{t}}=-11.7^{\circ}$. Both the incident and refracted spin-wave beams are on the same side of the interface normal, which indicates the occurrence of negative refraction. For the incident angles at $34.4^{\circ}<\theta_{\mathrm{i}}<90^{\circ}$, the refracted angles become positive and have the same sign with $\theta_{\mathrm{i}}$, recovering the normal refraction again, as shown in Fig. \ref{fig2}(f) with $\theta_{\mathrm{i}}=60^{\circ}$.

As we have shown, both the total reflection and the negative refraction can happen at the DMI interface for a certain range of the incident angles. Utilizing total reflection at the DMI interface, a spin-wave fiber or guide can be designed, analogous to the cases studied in Refs. \cite{Yu2016} and \cite{Mulkers2018}. It is worth noting that total reflection is not a unique feature at the DMI interface. In non-chiral ferromagnetic heterostructures with different material parameters such as the exchange constant, saturation magnetization, and thickness, it can happen as well \cite{JEONG2011,Xi2008,Gorobets1998,Vashkovskii1988}. In contrast, the negative refraction is exclusively due to the existence of a DMI step for the exchange spin wave and would disappear without the DMI. By measuring the incident and negative-refracted angles, one can determine the DMI constant based on the generalized Snell’s law [Eq. (\ref{eq7})].

One issue of our methods is how to excite the plane spin waves with the frequencies in the range of a few tens GHz in experiments. Very recently, it has been demonstrated that the excitation of the exchange spin waves with short wavelength (50 nm or shorter) and frequency up to 30 GHz can be realized by using the magnetization precession in periodic ferromagnetic nanowires to drive spin waves in a neighboring magnetic film \cite{Liu2018}.

\section{Three-magnon interactions arising in a magnetic strip}
In the above discussions, we focused on how to measure the DMI in ferromagnetic thin films of large scales. Due to the big laser spot size subjected to the diffraction limit of lights in the wavevector-resolved BLS, it is difficult to detect spin waves propagating in a rather narrow magnetic strip or nanowire. To address this problem, we propose a novel method to measure the DMI parameter in magnetic strip by analyzing the spectrum of spin waves outside the strip, which involves the nonlinear three-magnon processes.

\begin{figure}
  \centering
  \includegraphics[width=0.5\textwidth]{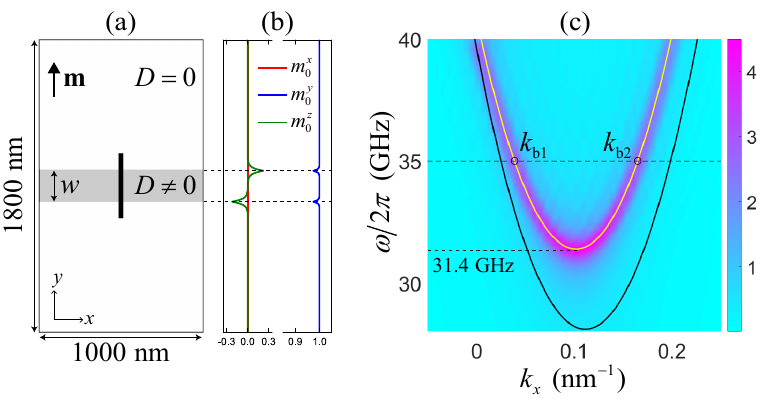}\\
  \caption{(a) The heterogeneous ultrathin film with a DMI strip (gray region) in the center. The width of the DMI strip is $w=50$ nm. The sinc-function field is applied in the regions of black bar. (b) Three components of the equilibrium magnetization $\mathbf{m}_{0}$ along longitudinal direction at $x=500$ nm. (c) FFT spectrum along the DMI strip center $y=900$ nm in (a). In (c), the black curve corresponds to the analytical formula Eq. (\ref{eq3}), while the yellow one represents the modified formula Eq. (\ref{eq8}) with the fitting parameter $\delta\approx25.1^{\circ}$. The wave vectors of spin waves with 35 GHz are $k_{\mathrm{b1}}=0.039$ $\mathrm{nm}^{-1}$ and $k_{\mathrm{b2}}=0.165$ $\mathrm{nm}^{-1}$, which correspond to the backward and forward spin waves, respectively.}\label{fig3}
\end{figure}
Firstly, we show that the three-magnon processes indeed can be induced by the DMI in uniform ferromagnets, even without the magnetic dipolar interaction. We start from the interfacial DMI Hamiltonian
\begin{equation}\label{9A1}
  \mathcal{H}_{\mathrm{DM}}=\frac{D}{M_{s}^{2}}\int{d\mathbf{r}}[M_{z}(\nabla\cdot\mathbf{M})-(\mathbf{M}\cdot\nabla)M_{z}],
\end{equation}
where the static magnetization $\mathbf{M}=(M_{x},M_{y},M_{z})$ lies in-plane and deviates from $\hat{x}$ direction with an arbitrary angle $\varphi$. We consider small oscillation of the magnetization over the ground state and represent $\mathbf{M}$ in the form $\mathbf{M}=\mathbf{M}_{0}+\mathbf{s}(\mathbf{r},t)$, where $\mathbf{M}_{0}$ is the background magnetization, and $\mathbf{s}$ corresponds to the small oscillations. We construct a new coordinate system for magnetization ($\mathbf{e}_{1},\mathbf{e}_{2},\mathbf{e}_{3}$) by rotating the coordinate system ($x,y,z$) around $\hat{z}$ over a angle $\varphi$, making $\mathbf{e}_{1}$ and $\mathbf{M}$ parallel. Expressing the magnetization in the rotated coordinate $(\mathbf{e}_{1},\mathbf{e}_{2},\mathbf{e}_{3})$ yields
\begin{equation}\label{M_e}
  \begin{split}
    M_{x} &= M_{1}\cos\varphi-M_{2}\sin\varphi, \\
    M_{y} &= M_{1}\sin\varphi+M_{2}\cos\varphi, \\
    M_{z} &= M_{3},
  \end{split}
\end{equation}
where $M_{1}=M_{0}+s_{1}$, $M_{2}=s_{2}$, and $M_{3}=s_{3}$. By Holstein-Primakoff transformation, we can express the magnetization in terms of boson operators ($a$ and $a^{+}$),
\begin{equation}\label{M_a}
  \begin{split}
    M_{1} &= 2\mu_{\mathrm{B}}(S-a^{+}a), \\
    M_{2} &= \mu_{\mathrm{B}}\sqrt{2S}(a+a^{+})-\frac{\mu_{\mathrm{B}}}{2\sqrt{2S}}(a^{+}aa+a^{+}a^{+}a), \\
    M_{3} &=-i\mu_{\mathrm{B}}\sqrt{2S}(a-a^{+})+\frac{i\mu_{\mathrm{B}}}{2\sqrt{2S}}(a^{+}aa-a^{+}a^{+}a),
  \end{split}
\end{equation}
where $S=M_{s}/(2\mu_{\mathrm{B}})$ is the spin of an atom with the Bohr magneton $\mu_{\mathrm{B}}$. Substituting (\ref{M_a}) and (\ref{M_e}) into (\ref{9A1}) and keeping the third-order terms of boson operators, we have
\begin{widetext}
\begin{equation}\label{A8}
  \begin{split}
    \mathcal{H}_{\mathrm{DM}}^{(3)} &= \frac{iDS^{-3/2}}{4\sqrt{2}}\int{d}\mathbf{r}\bigg\{\cos\varphi\left[4(a-a^{+})\frac{\partial}{\partial{x}}(a^{+}a)-4(a^{+}a)\frac{\partial}{\partial{x}}(a-a^{+})-\frac{\partial}{\partial{x}}(a^{+}aa-a^{+}a^{+}a)\right]\\ &+ \sin\varphi\left[4(a-a^{+})\frac{\partial}{\partial{y}}(a^{+}a)-4(a^{+}a)\frac{\partial}{\partial{y}}(a-a^{+})-\frac{\partial}{\partial{y}}(a^{+}aa-a^{+}a^{+}a)\right]\bigg\},
  \end{split}
\end{equation}
\end{widetext}
which is the three-magnon interaction Hamiltonian.

Next, we numerically examine the three-magnon processes arising in the DMI strip. To this end, we construct a heterogeneous ferromagnetic thin film with length 1800 nm, width 1000 nm, and thickness 1 nm, in which a DMI strip ($D=3.0$ $\mathrm{mJ/m^{2}}$) with the width $w=50$ nm locates in the center while the rest parts have no DMI, as shown in Fig. \ref{fig3}(a). To ascertain the spin-wave spectrum in the DMI strip, we apply a sinc-function field $\mathbf{h}(t)=h_{0}\sin[\omega_{\mathrm{H}}(t-t_{0})]/[\omega_{\mathrm{H}}(t-t_{0})]\hat{z}$ for 10 ns with $\mu_{0}h_{0}=0.01$ T, $\omega/2\pi=100$ GHz, and $t_{0}=1$ ns, over the black bar with volume $10\times400\times1$ $\mathrm{nm}^{3}$ shown in Fig. \ref{fig3}(a). In Fig. \ref{fig3}(c), the dispersion relation is obtained by performing the FFT of the spatiotemporal oscillation of the $z$-component magnetization ($\delta{m}_{z}$) over the lattices along the DMI strip center ($y=900$ nm) in Fig. \ref{fig3}(a). One can immediately see that the theoretical result (\ref{eq3}) [black curve shown in Fig. \ref{fig3}(c)] does not quite agree with the micromagnetic simulations. The reason for this deviation is attributed to the spin canting at the DMI interface \cite{Mulkers2017}, which is not a negligible effect any more because the tilting range is comparable with the strip width, as shown in Fig. \ref{fig3}(b). The largest tilting angle is obtained at the interface and is equal to $13.1^{\circ}$. The exact solution of the spin-wave spectrum on top of this strongly inhomogeneous magnetization texture is unlikely to obtain. However, to describe spin wave over this tilted ground state accurately enough, we assume that the DMI strip still has the uniform magnetization but deviating from the film plane with an angle $\delta$. The stabilization of this uniform tilting magnetization state requires an additional effective field $\mathbf{H}^{\prime}$ along $\hat{z}$ , which may originate from the DMI step \cite{Lee2017} or the pinning of magnetizations close to the interface. According to the equilibrium condition $\mathbf{m}\times\mathbf{H}_{\mathrm{eff}}=0$, we can get $\mathbf{H}^{\prime}=(H_{0}\tan\delta+M_{s}\sin\delta)\hat{z}$. Then, the effective field is given by
\begin{equation}\label{heff_DMIstrip}
  \mathbf{H}_{\mathrm{eff}}=\frac{2A}{\mu_{0}M_{s}}\nabla^{2}\mathbf{m}+\mathbf{H}_{\mathrm{DM}}
  +H_{0}\hat{y}-M_{s}m_{z}\hat{z}+\mathbf{H}^{\prime}.
\end{equation}
Substituting (\ref{heff_DMIstrip}) into the LLG equation (\ref{eq2}), the spin-wave dispersion relation in DMI strip can be calculated,
\begin{equation}\label{eq8}
  \begin{split}
     \omega= & \sqrt{(A^{\ast}\mathbf{k}^{2}+\omega_{\mathrm{H}}/\cos\delta)(A^{\ast}\mathbf{k}^{2}+\omega_{\mathrm{H}}/\cos\delta+\omega_{\mathrm{m}}\cos^{2}\delta)}\\       & -D^{\ast}k_{x}\cos\delta.
  \end{split}
\end{equation}

The simulated dispersion relation is well fitted by the above formula (\ref{eq8}) with $\delta\approx25.1^{\circ}$ [see the yellow curve in Fig. \ref{fig3}(c)]. However, we notice that the fitting parameter $\delta$ is twice as large as the actual tilting angle at the interface. This obvious disagreement is due to that approximating a strongly inhomogeneous magnetization texture by a globally tilted  magnetization state is too simplified. However, this approximation gives a good description of the dispersion relation of spin waves localized in the DMI strip. Interestingly, we find that the presence of the DMI reduces the spin-wave band gap from $\sqrt{\omega_{\mathrm{H}}(\omega_{\mathrm{H}}+\omega_{\mathrm{m}})}/2\pi\approx39.7$ GHz without the DMI to 31.4 GHz with $D=3.0$ $\mathrm{mJ/m^{2}}$, as shown in Fig. \ref{fig3}(c). In other words, spin waves with frequencies in the range (31.4, 39.7) GHz will be localized in the DMI strip. This motivates us to consider the three-magnon processes in the strip channel, while the present authors have considered a similar issue but in magnetic domain wall channels in Ref. \cite{Zhang2018}.

\begin{figure}
  \centering
  \includegraphics[width=0.35\textwidth]{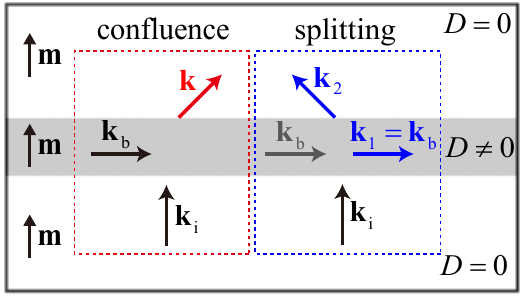}\\
  \caption{Schematic picture of nonlinear three-magnon processes in the DMI strip. In dashed red square, it shows the three-magnon confluence of $\mathbf{k}_{\mathrm{i}}$ and $\mathbf{k}_{\mathrm{b}}$ into $\mathbf{k}$. In dashed blue square, we plot the stimulated three-magnon splitting of $\mathbf{k}_{\mathrm{i}}$ into two modes $\mathbf{k}_{1}=\mathbf{k}_{\mathrm{b}}$ and $\mathbf{k}_{2}$, assisted by a localized magnon $\mathbf{k}_{\mathrm{b}}$ (gray arrow).}\label{fig4}
\end{figure}
In our strategy, we input a propagating spin wave ($\omega_{\mathrm{i}},\mathbf{k}_{\mathrm{i}}$) in the lower part of the heterogeneous films, to interact with the localized spin wave ($\omega_{\mathrm{b}},\mathbf{k}_{\mathrm{b}}$) bounded in the strip. In general, two kinds of three-magnon processes, i.e., confluence and splitting, can occur, as illustrated in Fig. \ref{fig4}. We first consider the three-magnon confluence. In this process, both the energy and the momentum parallel with the strip are conserved. Thus we have
\begin{equation}\label{eq9}
  \omega_{\mathbf{k}}=\omega_{\mathrm{i}}+\omega_{\mathrm{b}},\quad
  (\mathbf{k}-\mathbf{k}_{\mathrm{i}}-\mathbf{k}_{\mathrm{b}})\cdot \hat{x}=0.
\end{equation}
For a normal incident, i.e., $\mathbf{k}_{\mathrm{i}}=k_{\mathrm{i}}\hat{y}$, we obtain the wave vector of the the three-magnon confluence
\begin{equation}\label{confluence-k}
  \mathbf{k}=k_{\mathrm{b}}\hat{x}+k_{y}\hat{y},
\end{equation}
where $k_{y}\approx\sqrt{k_{\mathrm{i}}^{2}+C}$ with a positive constant $C=[\omega_{\mathrm{H}}/\cos\delta+(\omega_{\mathrm{m}}\cos^{2}\delta)/2-D^{\ast}k_{\mathrm{b}}\cos\delta]/A^{\ast}$.
For the three-magnon splitting process, the energy-momentum conservation gives rise to
\begin{equation}\label{eq10}
  \omega_{1}+\omega_{2}=\omega_{\mathrm{i}},\quad
  (\mathbf{k}_{1}+\mathbf{k}_{2}-\mathbf{k}_{\mathrm{i}})\cdot \hat{x}=0.
\end{equation}
In analogy to the case in Ref. \cite{Zhang2018}, the presence of spin waves localized in the DMI strip can trigger a stimulated splitting, implying $\mathbf{k}_{1}=\mathbf{k}_{\mathrm{b}}$ in (\ref{eq10}). We are also interested in the normal-incident case, so the wave vector of the three-magnon splitting can be determined as
\begin{equation}\label{splitting-k}
  \mathbf{k}_{2}=-k_{\mathrm{b}}\hat{x}+k_{2y}\hat{y},
\end{equation}
where $k_{2y}\approx\sqrt{k_{\mathrm{i}}^{2}-2k_{\mathrm{b}}^{2}-C}$. This indicates that the three-magnon splitting processes can only happen when $k_{\mathrm{i}}\geq\sqrt{2k_{\mathrm{b}}^{2}+C}$, which requires the frequency of the incoming spin waves higher than a critical value
\begin{equation}\label{wc_splitting}
  \omega_{\mathrm{i,c}}=A^{\ast}(2k_{\mathrm{b}}^{2}+C)+\omega_{\mathrm{H}}+\omega_{\mathrm{m}}/2.
\end{equation}

\begin{figure}
  \centering
  \includegraphics[width=0.5\textwidth]{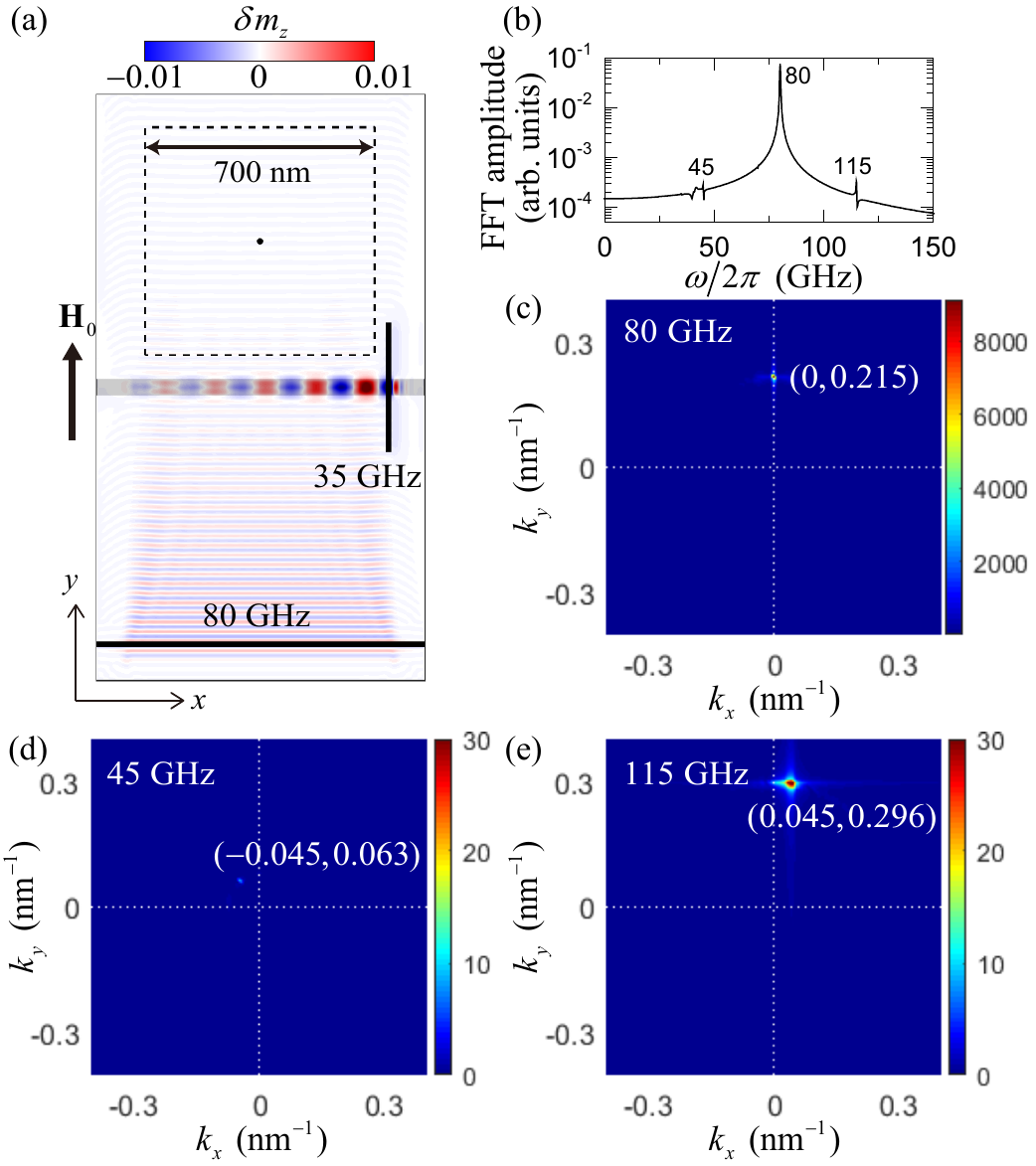}\\
  \caption{(a) Micromagnetic simulations of three-magnon processes. The incident and localized spin waves are excited at the lower part of the magnetic film (horizontal black bar) and the right side of the DMI strip (vertical black bar), respectively. (b) FFT spectrum at a single lattice [black dot in (a)]. (c)$\sim$(e) Spatial FFT spectra analyses for three peaks at (c) 80 GHz, (d) 45 GHz and (e) 115 GHz, observed in (b) where the incident spin-wave frequency is $\omega_{\mathrm{i}}/2\pi=80$ GHz and the localized spin-wave frequency is $\omega_{\mathrm{b}}/2\pi=35$ GHz. The FFT analysis is implemented over the region inside the dashed black square with the side length 700 nm in (a).}\label{fig5}
\end{figure}
Micromagnetic simulations are performed to verify the three-magnon processes arising in the DMI strip. We apply two sinusoidal monochromatic microwave fields simultaneously to excite the propagating spin waves ($\omega_{\mathrm{i}},\mathbf{k}_{\mathrm{i}}$) in the lower part of the magnetic film and the localized spin waves ($\omega_{\mathrm{b}},\mathbf{k}_{\mathrm{b}}$) in the DMI strip, respectively [see Fig. \ref{fig5}(a)]. Here, we consider $\omega_{\mathrm{i}}/2\pi=80$ GHz and $\omega_{\mathrm{b}}/2\pi=35$ GHz and focus on the normal incident case, i.e., $\mathbf{k}_{\mathrm{i}}\parallel{\hat{y}}$. Because of the conservation of both the energy and the momentum along the DMI strip ($\hat{x}$), the transmitted spin waves carry the information ($\omega_{\mathrm{i,b}},\mathbf{k}_{\mathrm{i,b}}$) from the incident and localized spin waves. This result is confirmed by the temporal FFT spectrum at a single cell [the black dot in Fig. \ref{fig5}(a)], which shows three peaks at 45 GHz, 80 GHz, and 115 GHz, as plotted in Fig. \ref{fig5}(b). The main peak of 80 GHz is from the incident spin wave excited at the lower part of the magnetic film. Two relatively weaker peaks at 45 GHz and 115 GHz are due to the three-magnon splitting and confluence processes, which satisfy the energy conservation $\omega_{\mathrm{k}}=\omega_{\mathrm{i}}\mp\omega_{\mathrm{b}}$, respectively. The wave vectors of spin waves for three frequency peaks can be obtained by the spatial FFT spectrum analysis over the region inside the dashed black square in Fig. \ref{fig5}(a). FFT results are shown in Figs. \ref{fig5}(c)$-$(e). The magnon wave vector at 80 GHz is $\mathbf{k}_{\mathrm{i}}=0.215\hat{y}$ in the unit of $\mathrm{nm}^{-1}$, which agrees with the dispersion relation Eq. (\ref{eq3}). While the magnon wave vectors at 45 GHz and 115 GHz are $\mathbf{k}=-0.045\hat{x}+0.063\hat{y}$ and $0.045\hat{x}+0.296\hat{y}$ in the units of $\mathrm{nm}^{-1}$, respectively. These values excellently agree with the wave vector formulas of the three-magnon splitting and confluence Eq. (\ref{splitting-k}) and (\ref{confluence-k}), respectively.
According to the conservation of momentum parallel with the DMI strip, the $x$-components of the wave vectors $\mathbf{k}$ of the transmitted spin waves for three-magnon confluence and stimulated splitting are $\mathbf{k}_{\mathrm{b}}$ and $-\mathbf{k}_{\mathrm{b}}$, respectively. Therefore, we can determine the wave vector of the localized spin wave in the DMI strip, $\mathbf{k}_{\mathrm{b}}=0.045\hat{x}$ in unit of $\mathrm{nm}^{-1}$, which is consistent with the direct FFT analysis in the strip.

\begin{figure}
  \centering
  \includegraphics[width=0.5\textwidth]{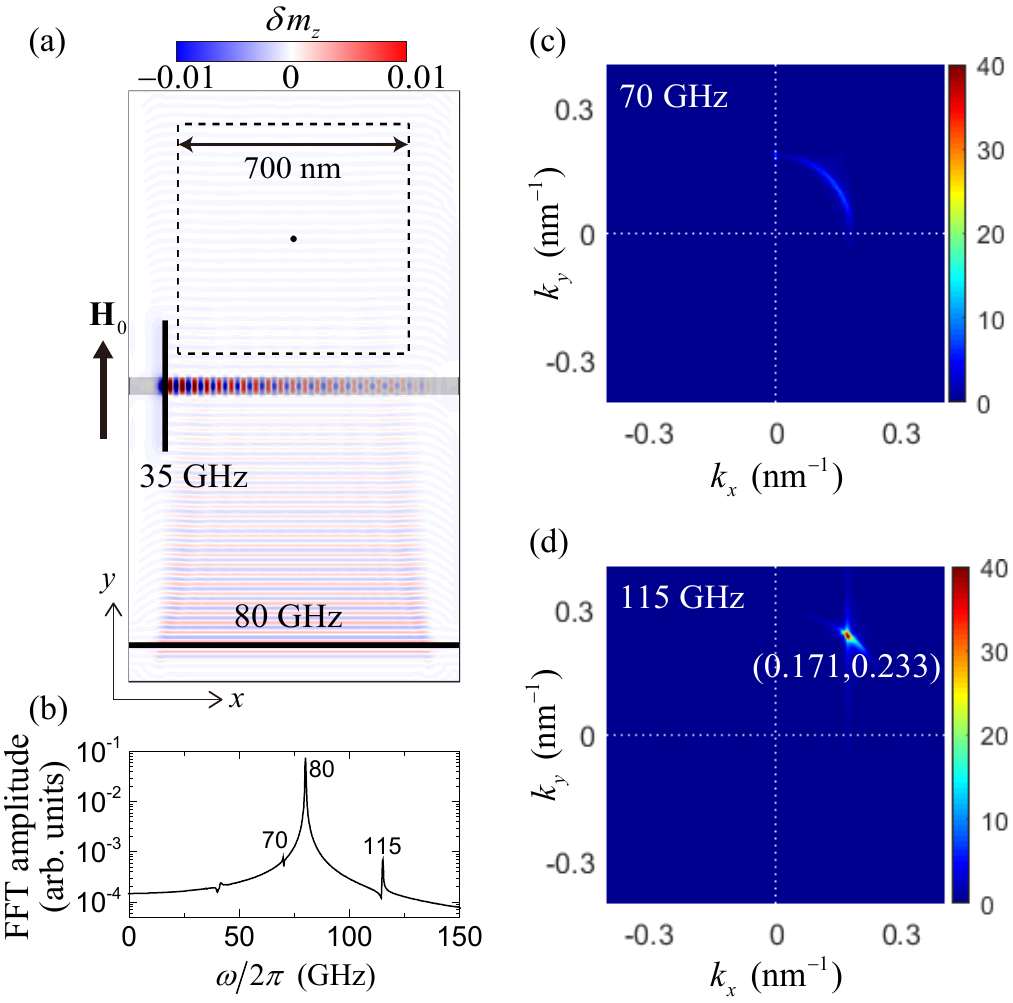}\\
  \caption{(a) Micromagnetic simulations of three-magnon processes with a different localized spin-wave mode. The incident and localized spin waves are excited at the lower part of the magnetic film (horizontal black bar) and the left side of the DMI strip (vertical black bar), respectively. (b) FFT spectrum at a single lattice cell [black dot in (a)]. Spatial FFT spectra analyses for the frequency $\omega/2\pi=70$ GHz (c) and $115$ GHz (d) are implemented over the region inside the dashed black square with the side length 700 nm in (a).}\label{fig6}
\end{figure}
Now, considering the inverse problem by assuming that both the DMI constant $D$  and the canting angle $\delta$ are two unknown parameters in Eq. (\ref{eq8}), only one group of ($\omega_{\mathrm{b}},k_{\mathrm{b}}$) is insufficient to determine them. We need another set of ($\omega_{\mathrm{b}},k_{\mathrm{b}}$) to completely quantify the DMI. To this end, we apply the same sinusoidal microwave field on the left side of the DMI strip, as shown in Fig. \ref{fig6}(a). A different microwave field on the same side also serves the same purpose (not shown). Although they have the same frequency, the localized spin waves excited at two sides of the DMI strip carry different wave vectors due to their non-reciprocal nature, as shown in Fig. \ref{fig5}(a) and Fig. \ref{fig6}(a). Temporal FFT spectrum analysis at a single cell [the black dot in Fig. \ref{fig6}(a)] shows two peaks at 80 and 115 GHz in Fig. \ref{fig6}(b), which are from the incident spin wave and the three-magnon confluence event discussed above. As compared with the FFT spectrum in Fig. \ref{fig5}(b), the frequency peak at 45 GHz disappears, but there is a very weak new peak at 70 GHz. The reason for the disappearance of 45 GHz peak is that the frequency of the incident spin wave is not high enough for generating the stimulated three-magnon splitting process. According to the criterion Eq. (\ref{wc_splitting}), the lowest incident frequency to generating the splitting process is $\omega_{\mathrm{i,c}}/2\pi\approx105.3$ GHz, when the localized spin waves with 35 GHz are excited at the left side. The appearance of 70 GHz peak is due to the frequency-doubling effect of the localized spin wave in the DMI strip. The spatial FFT spectrum of the peak at 70 GHz indicates that spin waves can propagate to a broad direction in the no-DMI regions [see the arc shown in Fig. \ref{fig6} (c)]. Based on the energy conservation in the three-magnon process, the frequency of the localized spin wave is the difference between two frequency peaks, i.e., $\omega_{\mathrm{b}}/2\pi=115-80=35$ GHz. The conservation of momentum parallel with the DMI strip indicates that the wave vector of the localized spin wave is $\mathbf{k}_{\mathrm{b}}=0.171\hat{x}$ in unit of $\mathrm{nm}^{-1}$, as shown in Fig. \ref{fig6}(d). Further, the wave vector $0.171$ $\mathrm{nm^{-1}}$ corresponds to a spin-wave wavelength 36.7 nm, which can be measured by an antenna as demonstrated in Ref. \cite{Liu2018}. Substituting the two sets ($\omega_{\mathrm{b}},k_{\mathrm{b}}$) into the dispersion relation (\ref{eq8}), and solving the following coupled equations
\begin{eqnarray}\label{eq11}
  \left\{
  \begin{aligned}
    \omega_{\mathrm{b}}&=\sqrt{(A^{\ast}k_{\mathrm{b1}}^{2}+\frac{\omega_{\mathrm{H}}}{\cos\delta})(A^{\ast}k_{\mathrm{b1}}^{2}+\frac{\omega_{\mathrm{H}}}{\cos\delta}+\omega_{\mathrm{m}}\cos^{2}\delta)}\\
    &-D^{\ast}k_{\mathrm{b1}}\cos\delta,\\
   \omega_{\mathrm{b}}&=\sqrt{(A^{\ast}k_{\mathrm{b2}}^{2}+\frac{\omega_{\mathrm{H}}}{\cos\delta})(A^{\ast}k_{\mathrm{b2}}^{2}+\frac{\omega_{\mathrm{H}}}{\cos\delta}+\omega_{\mathrm{m}}\cos^{2}\delta)}\\
    &-D^{\ast}k_{\mathrm{b2}}\cos\delta,
  \end{aligned}
  \right.
\end{eqnarray}
we obtain the DMI constant $D=3.4$ $\mathrm{mJ/m^{2}}$ and $\delta=32.7^{\circ}$, which is consistent with the input parameter $D=3.0$ $\mathrm{mJ/m^{2}}$ and the fitting $\delta=25.1^{\circ}$ obtained earlier. We also perform micromagnetic simulations with smaller DMI constants ($1.5\sim2.5$ $\mathrm{mJ/m^{2}}$), which are the typical values measured in experiments \cite{Stashkevich2015,Nembach2015}. The DMI constants obtained by solving Eq. (\ref{eq11}) are excellently consistent with the simulation parameters. These results suggest that the local DMI of a narrow magnetic strip can be accurately probed by non-locally detecting the spectra of both the incident and the transmitted spin waves involving in the nonlinear three-magnon processes.

\section{Conclusion}
To conclude, we systematically investigate the propagation, scattering, and interaction of spin waves in various ferromagnetic mediums and structures. In homogeneous ferromagnetic thin films, we predict a non-collinearity of two spin-wave beams with $\pm\mathbf{k}\parallel\mathbf{m}$, which solely comes from the DMI rather that the dipolar interaction. By measuring the angle between the two beams, one can determine the DMI parameter. We also consider a magnetic interface in the heterogeneous ultrathin films with different DMIs, and obtained the spin-wave Snell's law confirmed by micromagnetic simulations. Total reflection and negative refraction are observed at the DMI interface for certain incident angles. The total reflection induced by the DMI can be used to design spin-wave fiber with unidirectional transmission functionality. Negative refraction found here is exclusively induced by the DMI. These effects would provide an alternative approach to BLS for probing the DMI strength. Moreover, we propose a nonlocal scheme to measure the DMI parameter in a narrow ferromagnetic strip or nanowire by three-magnon processes, which is not accessible for the wavevector-resolved BLS due to the detection limit. Our results would be helpful to extend the present method for probing the DMI in experiments and for designing novel magnonic devices in the future.

\section{Acknowledgment}
We thank X.S. Wang, C. Wang, and Z.-X. Li for helpful discussions.
This work is supported by the National
Natural Science Foundation of China (Grants No. 11604041 and 11704060), the
National Key Research Development Program under Contract No. 2016YFA0300801,
and the National Thousand-Young-Talent Program of China.

\end{document}